\documentclass[preprint,aop]{imsart}

\usepackage{url}

\usepackage{pdflscape}
\usepackage{amsmath}
\usepackage{amssymb}
\usepackage{bbm}
\usepackage{amsthm}
\usepackage{graphicx}
\usepackage{xcolor}
\usepackage{multirow}
\usepackage{natbib,bm}
\usepackage{booktabs}
\begin{document}

\begin{frontmatter}
	
% Title of paper
\title{Integrating Biological Knowledge in Kernel-Based Analyses of Environmental Mixtures and Health}

% List of authors, with corresponding author marked by asterisk
\author{GLEN MCGEE$^\ast$\\%[4pt]
	% Author addresses
	\textit{Department of Statistics and Actuarial Science, 
		University of Waterloo,
		Waterloo, ON, Canada } % N2L3G1, 
	\\%[2pt]
	% E-mail address for correspondence
	{glen.mcgee@uwaterloo.ca} \\
	ANDER WILSON \\
		% Author addresses
		\textit{Department of Statistics, 
			Colorado State University
			Fort Collins, CO, USA } \\ %80523, USA
	BRENT A. COULL\\
		% Author addresses
		\textit{Department of Biostatistics, 
			Harvard T.H. Chan School of Public Health, 
			Boston, MA, USA} \\ % 02115, USA
		THOMAS F. WEBSTER\\
			% Author addresses
			\textit{Department of Environmental Health, 
				Boston University School of Public Health, 
				Boston, MA, USA} %02118, USA
}
%
%\author{ANDER WILSON\\[4pt]
%	% Author addresses
%	\textit{Department of Statistics, 
%		Colorado State University
%		Fort Collins, CO 80523, USA}
%	\\[2pt]
%	% E-mail address for correspondence
%	%{ander.wilson@colostate.edu}
%}
%
%\author{BRENT A. COULL\\[4pt]
%	% Author addresses
%	\textit{Department of Biostatistics, 
%		Harvard T.H. Chan School of Public Health, 
%		Boston, MA 02115, USA}
%	\\[2pt]
%	% E-mail address for correspondence
%%	{bcoull@hsph.harvard.edu}
%}
%
%\author{THOMAS F. WEBSTER\\[4pt]
%	% Author addresses
%	\textit{Department of Environmental Health, 
%		Boston University School of Public Health, 
%		Boston, MA 02118, USA}
%	\\[2pt]
%	% E-mail address for correspondence
%%	{twebster@bu.edu}
%}

%\author[A]{\fnms{Glen} \snm{McGee}\ead[label=e1]{glen.mcgee@uwaterloo.ca}},
%\author[B]{\fnms{Ander} \snm{Wilson}\ead[label=e2]{ander.wilson@colostate.edu}},
%\author[C]{\fnms{Brent A.} \snm{Coull}\ead[label=e3]{bcoull@hsph.harvard.edu}}
%\and
%\author[D]{\fnms{Thomas F.} \snm{Webster}\ead[label=e4]{twebster@bu.edu}}
%%%%%%%%%%%%%%%%%%%%%%%%%%%%%%%%%%%%%%%%%%%%%%%
%%% Addresses                                %%
%%%%%%%%%%%%%%%%%%%%%%%%%%%%%%%%%%%%%%%%%%%%%%%
%\address[A]{Department of Statistics and Actuarial Science, University of Waterloo,
%\printead{e1}}
%
%\address[B]{Department of Statistics, Colorado State University,
%\printead{e2}}
%
%\address[C]{Department of Biostatistics, Harvard T. H. Chan School of Public Health,
%	\printead{e3}}
%
%\address[D]{Department of Environmental Health, Boston University,
%	\printead{e4}}

% Running headers of paper:
\markboth%
% First field is the short list of authors
{G. McGee and others}
% Second field is the short title of the paper
{Integrating Knowledge in Mixtures Analyses}

\maketitle

% Add a footnote for the corresponding author if one has been
% identified in the author list
\footnotetext{To whom correspondence should be addressed.}

\begin{abstract}
{A key goal of environmental health research is to assess the risk posed by mixtures of pollutants. As epidemiologic studies of mixtures can be expensive to conduct, it behooves researchers to incorporate prior knowledge about mixtures into their analyses. This work extends the Bayesian multiple index model (BMIM), which assumes the exposure-response function is a non-parametric function of a set of linear combinations of pollutants formed with a set of exposure-specific weights. The framework is attractive because it combines the flexibility of response-surface methods with the interpretability of linear index models. 
We propose three strategies to incorporate prior toxicological knowledge into construction of indices in a BMIM: 
(a) constraining index weights, (b) structuring index weights by exposure transformations, and (c) placing informative priors on the index weights. We propose a novel prior specification that combines spike-and-slab variable selection with informative Dirichlet distribution based on relative potency factors often derived from previous toxicological studies. In simulations we show that the proposed priors improve inferences when prior information is correct and can protect against misspecification suffered by na\"ive toxicological models when prior information is incorrect. Moreover, different strategies may be mixed-and-matched for different indices to suit available information (or lack thereof). We demonstrate the proposed methods on an analysis of data from the National Health and Nutrition Examination Survey and incorporate prior information on relative chemical potencies obtained from toxic equivalency factors available in the literature. } %200 words 
%Keywords
{Informative priors; Environmental mixtures; Multiple index model  }
\end{abstract}

\end{frontmatter}

\section{Introduction}
\label{s:intro}

Understanding the risks of environmental pollutants has long been a public health priority, and many studies have investigated the association between a pollutant and a health outcome.  But humans are never exposed to a single pollutant in isolation; rather they are exposed to countless different pollutants. As such, environmental health research now routinely investigates the joint effects of \textit{mixtures} of exposures, and how pollutants contribute to them \citep{carlin2013unraveling,taylor2016statistical}. Studies of exposure to mixtures often rely on small datasets for which many chemical exposures have been assessed. It thus  behooves researchers to exploit any prior knowledge they may have about such mixtures \citep{reich2020integrative}. As \cite{thomas2007dissecting} argued,``by directly incorporating into our analyses information from other studies or allied fields—we can improve our ability to distinguish true causes of disease from noise and bias.''

In recent years, methods for analyzing environmental mixtures have proliferated, with proposed methods including: linear index models \citep{carrico2015characterization,keil2020quantile}, single and multiple index models \citep{wang2020singleindex,mcgee2021bayesian}, exposure-response surface methods \citep{bobb2015bayesian,bobb2018statistical,vieira2021assessing}, shrinkage and selection priors \citep{dunson2008shrinkage,herring2010nonparametric,antonelli2020estimating},  dimension reduction approaches like profile regression \citep{molitor2010bayesian,molitor2011identifying,molitor2014blood} and  Dirichlet process mixture models \citep{dunson2008bayesian, dunson2007bayesian}, among others  (for recent reviews, see \citealp{davalos2017current,hamra2018environmental,tanner2020environmental,joubert2022powering}). Despite these statistical advances, most existing approaches do not make use of available  knowledge about the toxicity of mixture components from the toxicological sciences.  \cite{reich2020integrative} proposed mechanisms for incorporating prior knowledge in variable selection and dimension reduction techniques, but we are unaware of any methods for integrating prior mixtures knowledge in non-linear models.

The recently proposed Bayesian multiple index model (BMIM; \citealp{mcgee2021bayesian}) is a compromise between the flexibility of response-surface methods like Bayesian kernel machine regression (BKMR; \citealp{bobb2015bayesian}) and the interpretability of linear index models (e.g., weighted quantile sum regression, WQS, \citealp{carrico2015characterization}; or quantile g-computation, QGC, \citealp{keil2020quantile}). 
The BMIM imposes structure on the BKMR framework by grouping mixture components  into linear indices, which are  weighted sums of a set of mixture components. 
The BMIM also adds flexibility to linear index models by allowing for interactions among indices and non-linear associations between the indices and a health endpoint.
Grouping mixture components into indices reduces the dimensionality  of the non-parametric estimation task and maintains interpretability by allowing  each index effect to be decomposed into component contributions via estimated weights. As we exploit here, these  weights provide an interpretable platform on which to incorporate prior biological knowledge.

Structuring exposures within linear indices is justified by toxicological models for multipollutant mixtures. Toxicologists often possess dose-response information for individual compounds and  characterize their joint response via some model of additivity (such as dose/concentration additivity). For example, the relative potency factor (RPF) model assumes that all components of the mixture have the same dose-response curve, differing only in potency---which is the amount of a compound needed to get a response 50\% of maximum \citep{howard2009generalized}. Specifically, components are assumed to act as substitutes for one another in proportion to their potency (relative to a reference compound), and the overall effects are estimated as a non-linear function of a sum of component doses/concentrations weighted by RPFs.  The best known examples of RPFs are called toxic equivalence factors (TEFs),  applied to dioxin-like compounds, using 2,3,7,8-TCDD as the reference compound \citep{van20062005}. RPFs are non-negative  weights  derived from experimental work with individual compounds.  In epidemiological studies, these non-negative weights are typically treated as known \textit{a priori} (e.g., \citealp{mitro2016cross,minguez2017longitudinal}). On one hand, RPFs may contain useful information about mixtures that could inform epidemiological analyses of mixtures; on the other,  there is uncertainty about whether RPFs are directly transportable to human populations.

In this paper we propose a suite of strategies for incorporating toxicological knowledge in environmental mixtures analyses via BMIMs. These include placing constraints on index weights, structuring weights (e.g., imposing effect rankings, or smoothness for temporally-ordered exposures, etc.) via exposure transformations, and adopting informative priors based on RPFs. 
We propose a novel prior specification that combines variable selection via a discrete mixture and an informative Dirichlet distribution. We show that when prior knowledge is correct, the proposed informative priors can improve inferences, and when prior information is incorrect, this approach can protect against mis-specification suffered by na\"ive methods that assume fixed weights. 
Software in the form of R code is available at {\tt github.com/glenmcgee/bsmim2}.

We briefly describe the case study of the association between a mixture of persistant organic pollutants and leukocyte telomere length in Section \ref{s:data}. In Section \ref{s:MIM}, we review the BMIM. In Section \ref{s:incorporating} we propose strategies for incorporating biological knowledge about exposure mixtures. In Section \ref{s:sim} we report on simulation studies investigating operating characteristics of the proposed approaches. In Section \ref{s:application} we apply the methods to the case study, and show how the proposed strategies can improve inferences. We conclude with a discussion in Section \ref{s:discussion}.

\section{Case Study: National Health And Nutrition Examination Survey}
\label{s:data}
We consider a case study of $N$=1003 people from the 2001-2002 cycle of the National Health and Nutrition Examination Survey (NHANES), in which \cite{mitro2016cross} first analyzed the association between a mixture of pollutants and the logarithm of leukocyte telomere length (LTL).  \cite{gibson2019overview} and \cite{mcgee2021bayesian} later reanalyzed the same sample, grouping 18 persistent organic pollutants into three  classes, containing: (1) eight non-dioxin-like PCBs; (2) two non-ortho PCBs; and (3) mono-ortho-PCB 118, four furans and four dioxins.  

\cite{mitro2016cross} adopted a toxic equivalent (TEQ) analysis, constructing a linear index based on a set TEFs assigned by the World Health Organization \citep{van20062005} and which were treated as known \textit{a priori}.  While these TEFs were based on experimental results, it is unclear whether they apply directly to the human population studied in NHANES. Moreover, none of this toxicologic information was leveraged by the more advanced mixture models considered by \cite{gibson2019overview} and \cite{mcgee2021bayesian}.  Here we consider a broader range of mechanisms for incorporating this  information.

\section{The  Bayesian Multiple Index Model (BMIM)}
\label{s:MIM}

Let $y_i$ be a continuous outcome of interest and  $\{x_{i1},\cdots,x_{iP}\}$ be a set of $P$ standardized exposures (i.e., mixture components) for the $i^{th}$ observation ($i=1,\cdots,n$).   Suppose $x_{i1},\cdots,x_{iP}$ are partitioned into $M$ ($M\in \{1,\dots,P\}$) mutually exclusive groups denoted  $\mathbf{x}_{im}=(x_{im1},\cdots,x_{imL_m})^T$ for $m=1,\dots,M$. Finally let $\textbf{z}_{i}$ be a vector of covariates with associated coefficient vector $\boldsymbol{\gamma}$. The Bayesian multiple index model (BMIM;  \citealp{mcgee2021bayesian}) is 
\begin{align}
y_i&=h^M\left(\mathbf{x}_{i1}^T \boldsymbol{\theta}_1,\cdots,\mathbf{x}_{iM}^T \boldsymbol{\theta}_M\right) +\mathbf{z}_i^T\boldsymbol{\gamma}+\epsilon_i,  ~~\epsilon_i\sim N(0,\sigma^2),\label{eqn:bmim} 
\end{align}
where $\boldsymbol{\theta}_m$ are $L_m$-vectors of index weights, and $h^M(\cdot): \mathbb{R}^M \to \mathbb{R}$ is an unknown and potentially non-linear function represented via a kernel function. 
A special case of BMIM is when $P=M$ and each component is in a separate index of size 1. In this case, BMIM is equivalent to BKMR.

 We assume $h^M(\cdot)$ exists in a space $\mathcal{H}_K$ defined by a positive semi-definite reproducing kernel $K:\mathbb{R}^M \times \mathbb{R}^M \to \mathbb{R}$.
The choice of kernel function $K(\cdot,\cdot)$ uniquely determines a set of basis functions \citep{cristianini2000introduction}.  Common choices include  the Gaussian kernel, $K(\mathbf{E},\mathbf{E'})=\exp\left[-\sum_{m=1}^M \rho_m (E_m-E'_m)^2\right],$ and the polynomial kernel of degree $d$, $K(\mathbf{E},\mathbf{E'})=\left[1+\sum_{m=1}^M \rho_m E_m E'_m \right]^d$, for $\mathbf{E}=(E_1,\cdots,E_M)$ and $\mathbf{E'}=(E'_1,\cdots,E'_M)$, and $\rho_m \geq 0$ are feature weights. With the index structure, $\mathbf{E}=(\mathbf{x}_{1}^T \boldsymbol{\theta}_1,\cdots,\mathbf{x}_{M}^T \boldsymbol{\theta}_M)$ and $\mathbf{E'}=(\mathbf{x^{'}}_{1}^T \boldsymbol{\theta}_1,\cdots,\mathbf{x^{'}}_{M}^T \boldsymbol{\theta}_M)$, and the Gaussian kernel can be written
$K(\mathbf{E},\mathbf{E'})=\exp\left[-\sum_{m=1}^M \rho_m \{(\mathbf{x}_{m}- \mathbf{x^{'}}_{m})^T \boldsymbol{\theta}_m\}^2\right].$ %\label{eqn:kern_gauss_BSMIM} %\\

Under a Kernel representation, model (\ref{eqn:bmim}) can be  written
\begin{align*}
y_i|h_i &\sim N(h_i+\mathbf{z}_i^T\boldsymbol{\gamma},\sigma^2), \\
(h_1,\cdots,h_N)^T&\sim N(\mathbf{0},\lambda^{-1}\sigma^2  \mathbf{K}),
\end{align*} 
where $\mathbf{K}$ is the kernel matrix with elements ${K}_{ij}=K(\mathbf{E}_i,\mathbf{E}_j)$ for  $\mathbf{E}_i=(\mathbf{x}_{1i}^T \boldsymbol{\theta}_1,\cdots,\mathbf{x}_{Mi}^T \boldsymbol{\theta}_M)$ and $\mathbf{E}_j=(\mathbf{x}_{1j}^T \boldsymbol{\theta}_1,\cdots,\mathbf{x}_{Mj}^T \boldsymbol{\theta}_M)$, and $\lambda>0$ is a tuning parameter that determines model complexity with small $\lambda$  favoring a more flexible model \citep{liu2007semiparametric}.

Without further constraints, the BMIM is over-parameterized, because $\rho_m$ and $\boldsymbol{\theta}_m$ are not independently identifiable. We thus impose  constraints to identify the sign and magnitude of $\rho_m$: (i) $\boldsymbol{1}_{L_m}^T\boldsymbol{\theta}_m\geq 0$, where $\boldsymbol{1}_{L_m}$ is the unit vector of length $L_m$, and  (ii) $\boldsymbol{\theta}_m^T\boldsymbol{\theta}_m =1$ for $m=1,\cdots,M.$  To circumvent constraint (ii)  in sampling from the posterior, we reparameterize the model in terms of $\boldsymbol{\theta}_m^*=\rho_m^{1/2}\boldsymbol{\theta}_m$ as in \cite{wilson2019kernel}.   Under this reparameterization, the Gaussian kernel can be written $K(\mathbf{E},\mathbf{E'})=\exp\left[-\sum_{m=1}^M  \{(\mathbf{x}_{m}- \mathbf{x^{'}}_{m})^T \boldsymbol{\theta}_m^*\}^2\right]$, and analogously for a polynomial kernel. We can then estimate the model in terms of $\boldsymbol{\theta}_m^*$ and  partition the posterior into $\rho_m=||\boldsymbol{\theta}_m^*||^{2}={\boldsymbol{\theta}_m^*}^T\boldsymbol{\theta}_m^*$ and $\boldsymbol{\theta}_m=||\boldsymbol{\theta}_m^*||^{-1}\boldsymbol{\theta}_m^*$.  In previous work we placed a weakly informative normal prior on ${\theta}_{ml}^*$ for $m=1,…,M$ and $l=1,\dots,L_m$, and further allowed for component-wise variable selection via a spike-and-slab prior:
 \begin{align*}
 {\theta^*_{ml}}|\nu_{ml}&\sim  \nu_{ml} \text{N}(0, \sigma_{\theta}^2) + (1-\nu_{ml})\delta_0, ~~\text{for $l=1,\cdots,L_m$  s.t. } \mathbf{1}_{L_m}^T\boldsymbol{\theta}_m^*\geq 0,
% \nu_{ml}&\sim \text{Bernoulli}(\pi), \\
% \pi&\sim \text{Beta}(a_0,b_0),
 \end{align*}
 where $\nu_{ml}\sim \text{Bernoulli}(\pi)$, $\pi\sim \text{Beta}(a_0,b_0)$, and $\delta_0$ is a point mass at zero. In this paper, we consider informative or constrained priors on either $\boldsymbol{\theta}_m$ or $\boldsymbol{\theta}_m^*$ that encode different forms of prior information that are often available in mixtures studies, which we will describe in Section \ref{s:incorporating}.

Finally we specify default priors for $\{\boldsymbol{\gamma},\sigma^2,\lambda\}$; see \cite{mcgee2021bayesian} for details. 

\subsection{Estimation and Interpretation}
We base estimation on the marginal likelihood of $\mathbf{y}=(y_1,\cdots,y_N)^T$ with respect to $\mathbf{h}=(h_1,\cdots,h_N)^T$,  $\mathbf{y}\sim N\left[\mathbf{Z}\boldsymbol{\gamma},\sigma^2(\mathbf{I}+\lambda^{-1} \mathbf{K})\right]$, 
where $\mathbf{Z}$ is the design matrix of covariates with $i^{th}$ row $\mathbf{z}^T_{i}$. Estimation proceeds via standard MCMC approaches (see \citealp{mcgee2021bayesian}).

%\subsection{Interpreting the Model}
To characterize the exposure-response surface, we estimate  $\mathbf{h}^{new}$ on a grid of $G$ new exposure levels, $\mathbf{E}^{new}_g$, $g=1,\dots, G$ (see \citealp{mcgee2021bayesian} for  details). In particular, we can describe index-wise response curves by varying the $m^{th}$ element over of a grid of index values---say, quantiles of the posterior means for $\boldsymbol{x}_{im}^T\boldsymbol{\theta}_m$---holding others constant.  These index-wise curves describe the shape of the response curve in relation to the entire $m^{th}$ index, treating weights $\boldsymbol{\theta}_m$ as fixed. The index weights $\boldsymbol{\theta}_m$ then quantify the relative contribution of each component, $x_{iml}$, to the effect of the $m^{th}$ index effect.  While the $m^{th}$ estimated index-wise curve ignores uncertainty in estimation of $\boldsymbol{\theta}_m$, we can quantify uncertainty for individual components via the posteriors  of the index weights themselves or via component-wise exposure-response curves, formed by predicting $\mathbf{h}^{new}$ for vector of indices $\mathbf{E}^{new}_g$ which vary over a grid of values for a single component.

\section{Incorporating Mixtures Knowledge}
\label{s:incorporating}
We consider three strategies for incorporating prior knowledge into a mixtures analysis with a BMIM: (a) constraints on the index weights, (b) linear transformations on the index weights, and (c) informative priors on the index weights. 
In each case, the chosen strategy applies to weights for a single index. In the presence of multiple indices, one can mix and match any of these strategies so that each index has a prior specification that matches the information available for that group of exposures, or has the default weakly informative prior when no prior information is available.

\subsection{Directional Homogeneity Constraints}
\label{ss:bmim_constrained}
It is often the case that exposures are believed to act in the same direction, such as when two compounds both act as agonists on the same receptor.  This assumption (sometimes known as directional homogeneity; \citealp{keil2020quantile}) is common in mixtures analyses: it underlies both WQS regression \citep{carrico2015characterization} and toxic equivalency analyses (use of multi-pollutant indices with TEFs treated as fixed, positive weights; e.g. \citealp{mitro2016cross}).

In the BMIM framework, we operationalize directional homogeneity on index $m$ by constraining  the implicit weights $\theta_{ml}^*\geq 0$ $\forall l=1,\dots,L_m$.  The constraint  $\theta_{ml}^*\geq 0$ is equivalent to the constraint on the index weights $\theta_{ml}\geq 0$, and  is guaranteed to satisfy the identifiability constraint $\boldsymbol{1_{L_m}}^T\boldsymbol{\theta}_m\geq 0$. Note that this constraint does not require the effects to be positive, simply that the components act in the same direction.

We impose this constraint by specifying the prior ${\theta^*_{ml}}\sim f_{\theta}(\theta^*)$, where  $f_{\theta}(\theta^*)$ is defined on the positive reals. 
We further implement variable selection via spike-and-slab prior:
\begin{align}
{\theta^*_{ml}}|\nu_{ml}&\sim  \nu_{ml} f_{\theta}(\theta^*)+ (1-\nu_{ml})\delta_0, ~~\text{for $l=1,\cdots,L_m$,  }\label{eqn:constraints}
\end{align}
where $\nu_{ml}\sim \text{Bernoulli}(\pi)$ and 
$\pi\sim \text{Beta}(a_0,b_0)$ as above.  Here we default to $f_{\theta}(\theta^*)\equiv\text{Gamma}(a_\theta,b_\theta)$, though any distribution on the positive real line could be used.

We can again decompose posterior samples of $\theta_{ml}^*$ to get estimates of  the L2-scale weights $\theta_{ml}$, which are subject to the standard L2 identifiability constraint, $\boldsymbol{\theta}_m^T\boldsymbol{\theta}_m =1$. Alternatively, the non-negativity constraint allows one to reparameterize yet again in terms of weights $w_{ml}=\theta_{ml}^*/\left(\sum_l \theta_{ml}^*\right)$. These weights $w_{ml}$ may be preferred due to their interpretation as proportions of the index effect (since $\sum_{l}w_{ml}=1$), as is typical of common linear index models such as quantile g-computation \citep{keil2020quantile} and WQS regression \citep{carrico2015characterization,colicino2019per}.  We can obtain posterior draws of these proportion-scale weights $w_{ml}$ by analogously deconvolving posterior draws of $\theta_{ml}^*$.

\subsection{Linear Transformations}
\label{ss:transformations}

\subsubsection{Ranked Weights}
\label{sss:bmim_rankorder}
In addition to assuming exposures act in the same direction, researchers often have prior knowledge about the relative ordering of their associations. Without loss of generality, assume exposures in the $m^{th}$ index are ordered from least to most potent: $0 \leq \theta_{m1}\leq \theta_{m2} \leq \cdots \leq \theta_{m{L_m}} $, or equivalently $0 \leq \theta^*_{m{1}}\leq \theta^*_{m{2}} \leq \cdots \leq \theta^*_{m{L_m}}$.  
To accommodate such an ordering, let $\beta_{m1},\dots,\beta_{m{L_m}} \geq 0$ such that $\beta_{m1}=\theta_{m1}^*$ and $\beta_{ml}=\theta_{ml}^*-\theta_{m(l-1)}^*$ for $l=2,\dots,L_m$. Equivalently, $\boldsymbol{\theta}_m^*=\boldsymbol{A}_m\boldsymbol{\beta}_m$ where $\boldsymbol{A}_m$ is an $L_m\times L_m$ lower triangular matrix of 1's. This yields:
$
\boldsymbol{x}_m^T\boldsymbol{\theta}_m^* =	\boldsymbol{x}_m^T\boldsymbol{A}_m\boldsymbol{\beta}_m={\boldsymbol{x}_m^*}{^T}\boldsymbol{\beta}_m,
$
where $\boldsymbol{x}_m^*={\boldsymbol{A}_m}^T\boldsymbol{x}_m$. In this case, we specify priors  for $\boldsymbol{\beta}_m$ directly.  As done for $\theta_{ml}^*$ in Section  \ref{ss:bmim_constrained}, we then assume a prior that induces non-negativity constraints on $\beta_{ml}$: $f_{\beta}(\beta_{ml})\equiv\text{Gamma}(a_{\beta},b_{\beta})$. Given a posterior sample of $\boldsymbol{\beta}_m$, we  obtain a posterior sample of $\boldsymbol{\theta}_m^*$ as $\boldsymbol{A}_m\boldsymbol{\beta}_m$, which also yields a posterior sample of the L2 weights $\boldsymbol{\theta}_m$, or the ``proportion'' weights $w_{ml}$ as defined in Section \ref{ss:bmim_constrained}.

In our software implementation of the model we also incorporate variable selection on the implicit weights $\boldsymbol{\beta}_{m}$ via spike and slab: 
\begin{align*}
{\beta_{ml}}|\nu_{ml}&\sim  \nu_{ml} f_{\beta}(\beta_{ml})+ (1-\nu_{ml})\delta_0, ~~\text{for $l=1,\cdots,L_m$  },
\end{align*}
with $\nu_{ml}\sim \text{Bernoulli}(\pi)$,
$\pi\sim \text{Beta}(a_0,b_0)$. 
Adopting spike-and-slab priors on the $\beta_{ml}$ has two implications for the weights $\theta^*_{ml}$. First, $\beta_{ml}=0$ when $\theta^*_{ml}=\theta^*_{m{(l-1)}}$, so  that variable selection on $\beta_{ml}$ encourages similar exposure weights to collapse to the same values. Second, there is an explicit spike-and-slab prior on the smallest weight, since $\theta^*_{m1}=\beta_{m1}$. Taken together, these imply that the spike-and-slab priors for $\beta_{ml}$ are able to shrink all the smallest weights to zero.

Fully ordering weights is powerful but restrictive. The same approach the same approach allows for partial orderings if we have  reliable prior information on  the relative orderings of only some of the effects.  For example, if interest focuses on an order contraint only on $k$ elements of an index, $0 \leq \theta_{m({L_m-k+1})} \leq \cdots \leq \theta_{m{L_m}}$, we leave the remaining $L_m - k$ elements unconstrained, $\theta_{m{1}},\dots,\theta_{m({L_m-k})}\geq 0$.  In the case where $k=2$ and $L_m=4$, this can be achieved with the transformation matrix ${\boldsymbol{A}_m}=\left[\begin{array}{cccccc}
1 & 0 &0 & 0  \\
0 & 1 & 0 & 0  \\
0 & 0  & 1 & 0 \\
0 & 0  & 1 & 1  \\
\end{array}\right].$ Alternatively, if knowledge exists on the relative ordering of the two weakest elements but there is uncertainty about which other component was most potent, one could set 
${\boldsymbol{A}_m}=\left[\begin{array}{cccccc}
 1 & 0 &    	0 & 0   \\
 1 & 1  &   	0 & 0  \\
 1 & 1    &	1 & 0  \\
 1 & 1  &	0 & 1  \\
\end{array}\right].$ 

\subsubsection{Smoothly Varying Weights}
\label{sss:smooth}
Rather than an index of separate mixture components assumed to act similarly on the outcome, 
interest often focuses on exposures to a single  component measured longitudinally . Let $x_{im}(t)$ be exposure to the $m^{th}$ mixture component for individual $i$ at time $t$, for time points $t=1,\dots,T$. \cite{wilson2019kernel} take a functional approach, defining a weighted exposure (index) $E_{im}=\int x_{im}(t) \theta(t) dt$, where $\theta(t)$ is a weight function that varies smoothly over time $t$. Both the exposure and the weight function are represented via basis function expansions: $x_{im}(t)=\sum_{j=1}^{J_{m}} \xi_{imj} \psi_{mj}(t) $ and $\theta(t)=\sum_{j=1}^{J_{m}} \beta_{mj} \psi_{mj}(t) $, where $\{\psi_{mj}(t)\}_{j=1}^{J_m}$ is a common orthonormal basis used in both expansions, and $\boldsymbol{\xi}_{im}=( \xi_{im1},\dots, \xi_{im{J_m}})^T$ and $\boldsymbol{\beta}_{m}=( \beta_{m1},\dots, \beta_{m{J_m}})^T$ are coefficient vectors.  The weighted exposure index  can then be written as $E_{im}={\mathbf{x^*}_{im}}^{T} \boldsymbol{\beta}_m,$ where $\mathbf{x}^*_{im}=\boldsymbol{\Psi}_m^T \mathbf{x}_{im}$.
As such we impose smoothness by  selecting an appropriate orthonormal basis, pre-transforming the exposures, and proceeding as usual by specifying default priors on $\boldsymbol{\beta}_m $ as for $\boldsymbol{\theta}_m$. 
Ultimately, this allows analysts to incorporate knowledge of the structure of exposure---the fact that the resulting effects are likely to vary smoothly over time---within a BMIM framework, and the estimated functional weight further allows one to investigate windows of susceptibility during which outcomes are most affected by a time-varying exposure (see \citealp{wilson2019kernel} for details). 
This approach is powerful in that it allows the effects of distributed lags and other structured indices to vary by other groups of co-exposures, such as other classes of chemicals, psychosocial stressors, or nutrients, among others.

\subsection{RPF-Centered Informative Priors}
\label{ss:informative}

\subsubsection{Targeted Dirichlet}
\label{sss:bmim_dirichlet}
As mentioned in Section \ref{ss:bmim_constrained}, when the $\theta_{ml}^*$ are constrained to be non-negative, it is advantageous to re-parameterize the model via $\theta_{ml}^*=\rho'^{1/2}_mw_{ml}\geq 0$  such that  $\sum_{l}w_{ml}=1$. Such a parameterization presents a mechanism for specifying informative priors. One convenient choice is a Dirichlet prior:
\begin{align*}
(w_{m1},\cdots,w_{mL_m}) &\sim  \text{Dirichlet}(\alpha_{m1},\dots,\alpha_{mL_m}).
\end{align*}
We separately specify a prior $f_r(\cdot)$ on the non-negative reals for $\rho'^{1/2}_m$; here we adopt a
$\text{Gamma}(a_{\rho},b_{\rho})$ as a default prior. 
Via a change of variable this induces the following prior for $\boldsymbol{\theta}^*_m$:
\begin{align*}
f(\theta^*_{1},...,\theta^*_{L_m})=\Gamma( \sum_{l}^{L_m}\alpha_l )\prod_{l}^{L_m}\frac{(\theta^*_{ml})^{\alpha_l -1}}{ \Gamma(\alpha_l)}\left(\sum_l^{L_m} \theta^*_{ml}\right)^{1-\sum\alpha_{l} } f_r\left(\sum\theta^*_{ml} \right)
\end{align*}
constrained by $\theta_{ml}^*\geq 0$. Hence, we can use this form of the prior on $\boldsymbol{\theta}^*$ to sample directly from the posterior of $\boldsymbol\theta^*$. We then decompose the posterior sample of $\boldsymbol\theta^*$ into samples of $\boldsymbol{w}_m$ and $\rho_m$.

 The main appeal of this Dirichlet specification is that it provides a means of encoding prior knowledge about the relative contributions of each exposure in a meaningful way via the relative sizes of $\alpha_{ml}$. This is especially useful in settings where researchers have previously used multipollutant indices based on fixed index weights like RPFs. The best known example are TEFs, which are fixed weights often used to construct a multipollutant index ${\Phi}=\sum_l a_{ml} x_{ml}$ (e.g., \citealp{mitro2016cross}). Without loss of generality, assume weights $a_{ml}$ have been scaled in order to sum to 1. Such weights $a_{ml}$ are typically  derived from theoretical or experimental results, which may not apply directly to human populations. Rather than using these values directly---or discarding them entirely---we specify the prior hyperparameters so that the prior  means for the proportion weights $w_{ml}$ are proportional  to the RPFs. We set $\alpha_{ml}=c a_{ml} $ for $l=1,...,L_m$, which implies
\begin{align*}
E[w_{ml}]=a_{ml},~~~~~~~Var[w_{ml}]=\frac{1}{1+c} \left(a_{ml} \right) \left(1-a_{ml}\right);
\end{align*}
hence $c$ can be used to tune the desired level of uncertainty around the experimental weights. In particular, larger values of $c$ result in a stronger, more informative, prior.

This informative Dirichlet prior can be viewed as a less rigid form of the order constraint.  Ordering weights as in Section \ref{sss:bmim_rankorder} can be overly restrictive at times, forcing weights to adhere to a strict hierarchy. Instead, this informative prior strategy relaxes this strong ordering assumption by using the informative Dirichlet-prior specification, choosing hyperparmaters $\alpha_{ml}$ to reflect the same hypothesized ranking of exposures without imposing rigid constraints. This allows estimates to deviate from the hypothesized ranking somewhat, thus protecting against misspecification.

A special case is use of a flat $\text{Dirichlet}(c,\dots,c)$ prior that puts prior mass on all $L_m$ weights being equal to $1/L_m$. When $c$ is large, this is an informative prior that encourages the index to be proportional to the average exposure value. 

\subsubsection{Targeted Dirichlet  with Component Selection}
\label{sss:dirichlet_selection}
The Dirichlet prior specification is useful when one has prior knowledge of index weights based on previous research, but it does not incorporate variable selection.   Nevertheless, there may be cases where potency information is available and variable selection is of interest. 

To that end, we combine the basic building blocks described above to construct a novel spike-and-slab prior formulation.  First, we leverage the relationship between Dirichlet and gamma distributions: $\theta^*_{ml} \overset{ind}{\sim} \text{Gamma}(\alpha_{ml},b_\theta)$ for $l=1,\dots,, L_m$, implies $(w_{m1},\cdots,w_{mL_m}) \sim  \text{Dirichlet}(\alpha_{m1},\dots,\alpha_{mL_m})$.  Second, we  incorporate variable selection as in Section \ref{ss:bmim_constrained}, replacing (\ref{eqn:constraints}) with
\begin{align}
{\theta^*_{ml}}|\nu_{ml}&\sim \nu_{ml} f_{\theta_{ml}}(\theta^*_{ml})+ (1-\nu_{ml})\delta_0, ~~\text{for $l=1,\cdots,L_m$,  } 
\end{align}
where $f_{\theta_{ml}}(\theta^*_{ml})=\text{Gamma}(\alpha_{ml},b_\theta)$. We then select $\alpha_{ml}$ as above. This allows one to incorporate prior information via a Dirichlet ``slab'' while simultaneously allowing for component selection. Note that $a_{ml}$ is no longer the prior mean for $w_{ml}$ because of the discrete mixture. Rather it is the prior mean for the ``slab'' component; that is, conditional on being selected ($w_{ml}\neq 0$), the prior mean of $w_{ml}$ is proportional to $a_{ml}$. 

\begin{figure}[htbp!]
	\centering
	\includegraphics[width=0.7\linewidth]{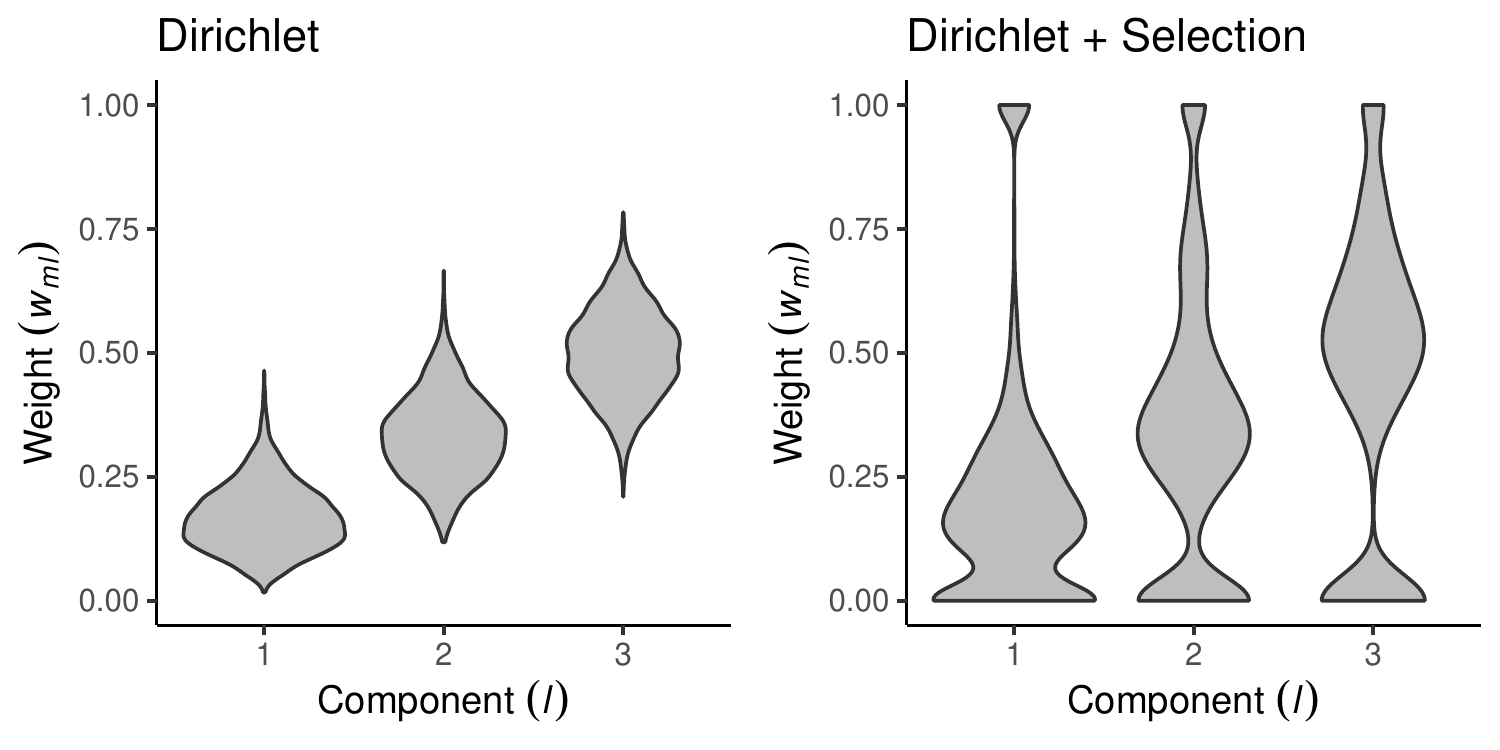}
	\caption{Informative prior distributions on component weights in an index with $L_m=3$ components. Left panel depicts  a $\text{Dirichlet}(5,10,15)$ distribution for $(w_{m1},w_{m2},w_{m3})$; right panel depicts the same Dirichlet slab but incorporates component selection with a prior inclusion probability of 75\%.}
	\label{fig:violininfprior}
\end{figure}

We visualize this prior for a mixture of three components in Figure \ref{fig:violininfprior}. The left panel depicts a $\text{Dirichlet}(5,10,15)$ prior on the proportion weights $(w_{m1},w_{m2},w_{m3})$; the right panel depicts the novel prior which combines the same $\text{Dirichlet}(5,10,15)$ with componentwise variable selection, with a prior inclusion probability $P(w_{ml}>0)=0.75$. Naturally, variable selection adds a point mass at 0, and as a result the distributions become slightly right-skewed and result in a small point mass at 1, corresponding to all other components being excluded ($w_{ml}=0$).

\section{Simulations}
\label{s:sim}

We conducted several simulation studies to investigate the impacts of the proposed strategies for incorporating prior knowledge into a BMIM analysis, both when prior knowledge is correct (Simulation A) and when it is mis-specified (Simulations B \& C).

\subsection{Simulation Setup}
\label{ss:setup}
We generated $R$=500 datasets of $n$=200 observations as follows. Using real exposure and covariate data from the NHANES sample (described in Section \ref{s:data}), we generated outcomes as
\begin{align*}
y_i&\sim N\{h(\mathbf{x}_{i1}^T \boldsymbol{w_1}  )+\mathbf{z}_i^T\boldsymbol{\gamma},\sigma^2\},
\end{align*}
where $\mathbf{x}_{i1}$ is a vector of $p=8$ pollutants, and $\mathbf{z}_i$ included age (standardized), age$^2$, male (0,1), and indicators of BMI category (25–-29.9; 30+). We set $\boldsymbol{\gamma}=[-0.43,0.00,-0.25,0.12,0.08]^T$,  $\sigma=0.5$, and %$h(x)=5 f(x/1.4)$ where $f$ is the $N(0,1)$ density  
$h(x)$ is a non-linear exposure response function (Figure A in the Supplementary Material).

In Simulation A, we explored the effect of incorporating (correct) prior knowledge about a mixture. We set $\boldsymbol{w}^A=[0.50, 0.25, 0.10, 0.05, 0.05, 0.02, 0.02, 0.01 ]^T$. To each dataset, we  fit six  single index models that each incorporate correct prior information about the weights $\boldsymbol{w}$ in different ways. Specifically, we fit: (i) an unconstrained model with variable selection; (ii) a constrained  model (with non-negative weights, i.e. directional homogeneity) with variable selection; (iii) a targeted Dirchlet model, with prior mass centered around the true weights, $\boldsymbol{w}^A$; (iv) a targeted Dirichlet model with variable selection; (v) a rank-ordered model that assumes $w_{j}\geq w_k$ $\forall j < k$; and (vi) a TEQ model that takes $\boldsymbol{w}^A$ as fixed and correct. Models (i)-(vi) are ordered by how much information they incorporate: model (i) incorporates no prior knowledge about the weights, whereas model (vi) assumes they are completely known a priori.  All models  assumed a Gaussian kernel. We also fit a full BKMR model for comparison.% (see Supplementary Material).

In Simulations B and C, we explored the impact of incorporating incorrect prior knowledge about a mixture. In Simulation B, we generated data according to {$\boldsymbol{w}^B=[0.10, 0.25, 0.50, 0.05, 0.05, 0.02, 0.02, 0.01 ]^T$} and fit models with the previously described priors that incorporate information about $\boldsymbol{w}^A$. This allows one to  investigate  the impact of incorporating incorect information about the relative weights of mixture components. In particular, models (v) and (vi) are strictly mis-specified in Scenario B (the true weights are not in the parameter space of the prior), and models (iii) and (iv) have priors centered around incorrect weights.  In Simulation C, we investigated the impact of incorrectly assuming directional homogeneity by generating data under $\boldsymbol{w}^C=[0.50, -0.25, 0.10, 0.05, 0.05, 0.02, 0.02, 0.01 ]^T$; here all models except for the unconstrained model (i) and BKMR were strictly mis-specified. 

In each scenario, we computed mean squared error (MSE), 95\% credible interval (CI) coverage and width for estimates of: (a) the exposure response surface ($h^{\texttt{new}}$)
 for a hold-out set of 200 real exposure vectors, and (b) component-wise curves (the exposure response resulting from varying a single exposure between its 25th and and 75th percentile, holding others at their medians), averaged over a grid of exposure values and further averaged over the eight  exposures. 
Throughout, we report relative MSEs and  widths by dividing by the corresponding values for the unconstrained model (i), with values less than 1 indicating better performance.
Finally, we compare distributions of posterior means for the weights $\boldsymbol{w}$ themselves (see Supplementary Material).

\subsection{Results}
\label{ss:results}
We summarize results of Simulations A and B in Table \ref{tab:simA} (see Table A1 in the Supplementary Material for Simulation C.). Naturally, when prior knowledge was based on correct weights (Simulation A), the models that incorporated more prior knowledge tended to have lower MSE and CI widths. As an upper bound on performance, the TEQ approach, which assumed weights were known a priori, resulted in MSE reduction of 42\% (MSE ratio 0.58) in estimating new $h$ values on a hold-out sample. The targeted Dirichlet models with selection (iii) and without selection (iv) also achieved large reductions in MSE (29\% and 35\%, respectively). In general, reductions in MSE and CI widths were even more pronounced for component-wise curves, though these represent somewhat artificial estimands. Estimates of the surface for a hold-out sample better reflect the full data generating mechanism.

When incorrect knowledge about relative weights was incorporated (Simulation B), the TEQ approach performed worst, with an increase in MSE of  118\% (MSE ratio of 2.18) compared to the unconstrained model (i), and the rank-ordered approach yielding an increase in MSE of 10\%. By contrast the informative Dirichlet models still performed well. Although they performed marginally worse than the constrained approach (ii) in MSE, they achieved lower average interval width. Moreover, both still performed better than the unconstrained model (i). While the informative priors were centered around incorrect values, they still allowed for uncertainty around those values, whereas the TEQ approach and rank-ordered approach were strictly mis-specified. 

Collectively, these results demonstrate that the targeted Dirichlet approach (with or without variable selection) can be an effective way to incorporate prior knowledge. When prior knowledge is correct it can lead to important gains in accuracy, and when prior knowledge is incorrect it does not pay as high a penalty as a model that \textit{assumes} incorrect weights.

\section{Analysis of NHANES Case Study}
\label{s:application}

We applied the proposed methods on the NHANES sample ($N$=1003). In particular, we are  interested in incorporating toxicological information about the third class of pollutants (containing mono-ortho-PCB 118, dioxins 1--3, and furans 1--4). We considered both a single index model containing only the third pollutant class, and  a 3-index model with each class defining an index.

We fit: (i) an unconstrained BMIM with the default weakly informative priors, (ii) a constrained BMIM that maintains directional homogeneity, (iii) a targeted Dirichlet prior analysis with prior means centered at the TEFs given in \cite{mitro2016cross}, (iv) a targeted Dirichlet prior that further incorporates variable selection, and (v) a TEQ analysis in which the index weights are treated as known and equal to the (scaled) TEFs. In the single index models, this information applies to the entire mixture ($P=8$), whereas in the three-index model, prior information is incorporated only for the third class of pollutants, and we leave the other indices unconstrained to reflect a lack of prior information.  All models were adjusted for age (linear and quadratic), sex and BMI category ($<$25, 25–30, $\geq$30), and we used a Gaussian kernel throughout.

We visualize the induced priors for the weights on the proportion scale ($w_{ij}$) for the third class of pollutants in the first row of Figure \ref{fig:gridplots}.  The constrained approach is weakly informative in that it reflects no prior information about the relative contributions of the mixture components; the Dirichlet approach, by contrast, centers mass around the TEFs, while still allowing for uncertainty around them. Results for the unconstrained approach are included in the supplementary material. % since weights are only available on the L2 scale.

\subsection{Results}
\subsubsection{Single Index Models}

\begin{figure}[htbp!]
	\centering
	\includegraphics[width=1.0\linewidth]{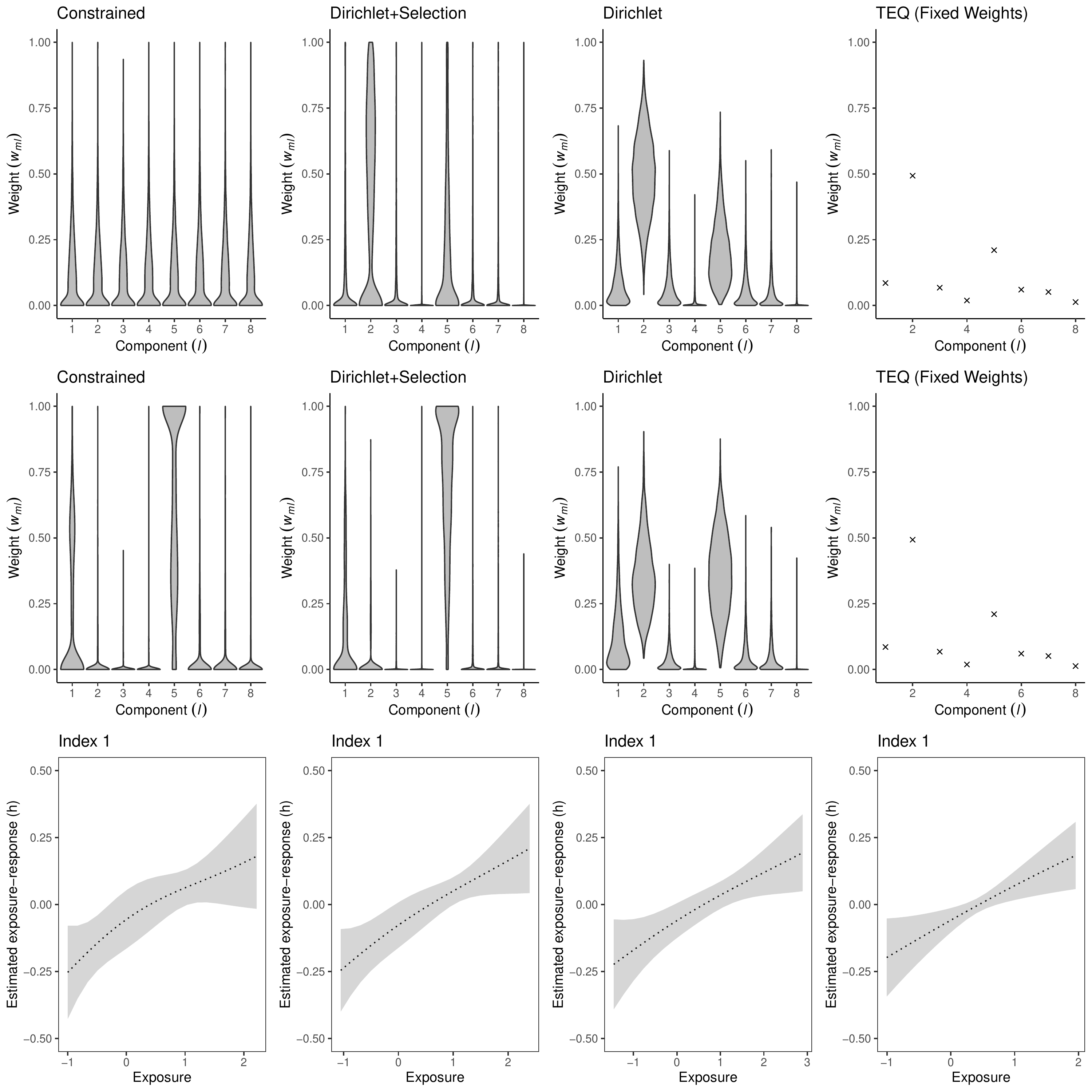}
	\caption{Results of the NHANES analysis using single index models to model the third class of pollutants. First row shows prior distributions for the `proportion' weights, $w$; second row shows the corresponding posterior distributions. Third row shows estimated indexwise curves. First column shows results for the non-informative constrained prior, second column shows the targeted Dirichlet approach with variable selection, the third column shows the targeted Dirichlet without selection, the fourth column shows the TEQ approach with fixed weights. Exposure components 1--8 correspond to PCB 188, dioxins 1--3, and furans 1--4.}
	\label{fig:gridplots}
\end{figure}

In the second row of Figure \ref{fig:gridplots}, we show the corresponding posterior distributions for proportion weights under each approach. In the constrained analysis,  Furan 1  dominated the index, with nearly all posterior weight being assigned to it. By contrast, under the targeted Dirichlet approach, the posterior weights are large for Furan 1 as well as Dioxin 1, which received the most prior mass. Interestingly, incorporating variable selection into the targeted Dirichlet approach yielded results more similar to those of the constrained approach than those of the standard targeted Dirichlet approach. Reducing the degree of variable selection (e.g. prior inclusion probabilities of 0.8 rather than 0.5) yielded similar results. 

The third row of Figure \ref{fig:gridplots} depicts estimated index-wise curves, corresponding to contrasts between the multipollutant index set to its $q^{th}$ quantile and its median. Unsurprisingly, the more informative the priors, the lower the posterior uncertainty, and the TEQ approach resulted in the narrowest credible intervals. Assuming the TEFs are correctly specified, the TEQ curve represents an upper bound on performance: we could not hope to do better than this, as it assumes no uncertainty related to the weights. Ultimately the informative Dirichlet approaches reduced interval widths relative to the constrained approach. This is in spite of the fact that the index-wise curves treat weights to be fixed at their posterior means, so we would not expect as dramatic gains here as in component-wise curves.

Estimated index-wise curves were remarkably similar across approaches, despite the estimated weights being very different. This is in part due to the high correlation between mixture components. The less informative priors placed most posterior mass on Furan 1, whereas the TEQ approach weighted Dioxin 1 most heavily, with the targeted Dirichlet lying somewhere in between. Because these two mixture components are so highly correlated (correlation of 0.73), they are somewhat interchangeable. %: intuitively, for  $u_1$ and $u_2$ that are highly correlated, $u_2$ and $(0.5 u_1 +0.5 u_2)$ are also highly correlated. 
Nevertheless, if the TEFs are incorrect and the true weights are better reflected by the data-driven posteriors, then the TEQ results are less transportable. That is, they would fare much worse in samples where Furan 1 and Dioxin 1 are less correlated. Even in the NHANES sample, the Dirichlet approach with variable selection estimated a mean outcome of 0.06 (95\% CI [-0.09, 0.21]) at the 75th percentile of Dioxin 1 and the 25th percentile of Furan 1 and a much higher 0.21  (95\% CI [0.07, 0.35]) at the 25th percentile of Dioxin 1 and the 75th percentile of Furan 1. The TEQ approach, by contrast, yielded estimates of 0.16 (95\% CI [0.03, 0.29]) and 0.12 (95\% CI [-0.02, 0.25]), respectively.

We compared fit by root mean squared error (RMSE) via 4-fold cross validation. While all models had similar fits in terms of RMSE (704---716), the targeted Dirichlet model and the TEQ model had the lowest (704 and 705), beating the targeted Dirichlet without selection (708).

%\vspace{-0.5cm}
\subsubsection{Multiple Index Models}

Results of the multiple index analysis are largely similar to those shown for the single index analysis, see Figure B1 in the Supplementary Material.

\begin{figure}[htbp!]
	\centering
	\includegraphics[width=1.1\linewidth]{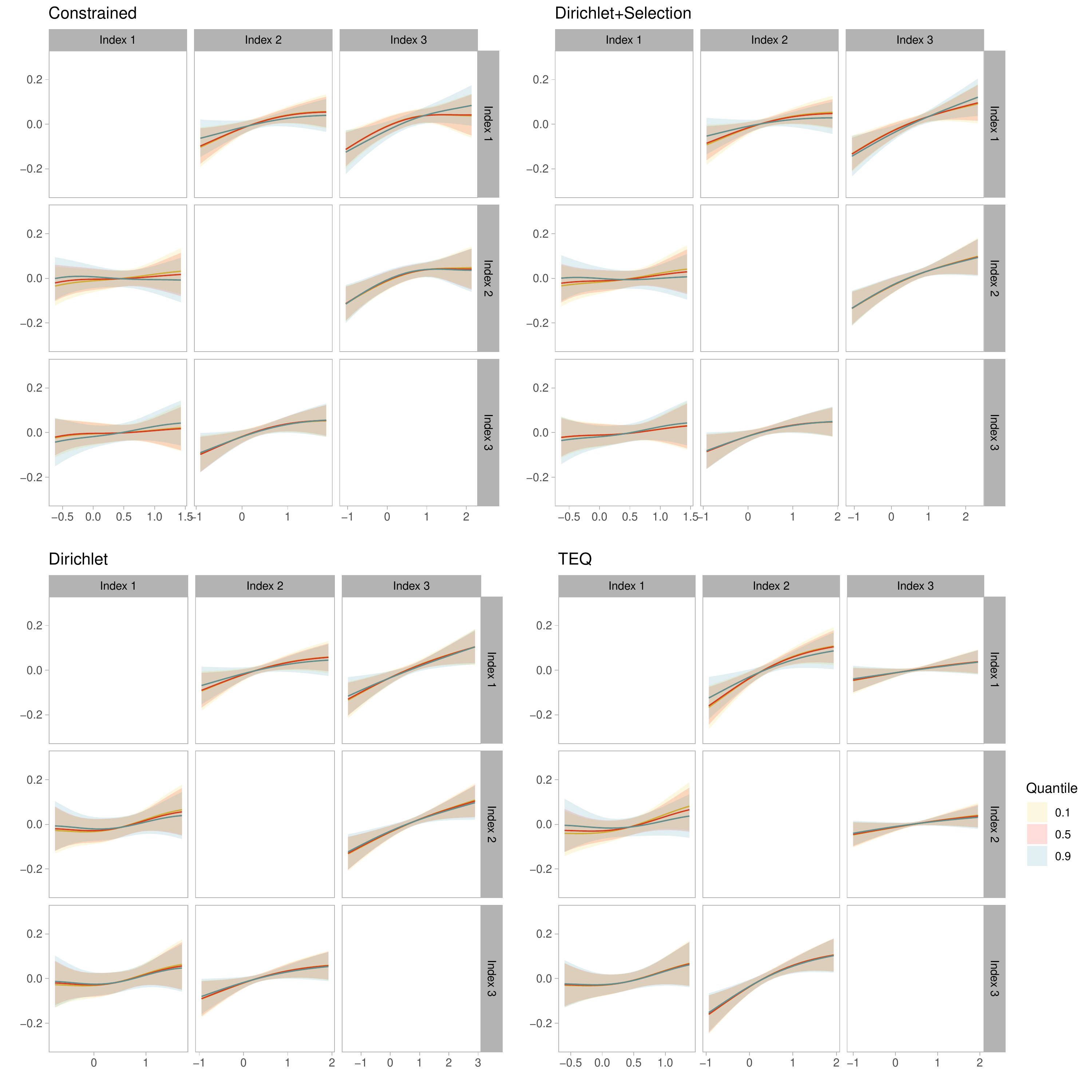}
	\caption{Interactions in the NHANES analysis using 3-index models.  Shown are estimated indexwise curves, holding another index at its 10th, 50th, and 90th percentile (and the other index at its median). Curves are centered to ignore additive effects; changes in slope or shape indicate interaction. Bands indicate approximate 50\% credible intervals.
	}
	\label{fig:grid_plots_inter_centered}
\end{figure}

A key advantage of the multiple index framework is the ability to investigate non-additive interactions among indices. In Figure \ref{fig:grid_plots_inter_centered} we plot estimated index-wise curves, holding another index at its 10th, 50th and 90th percentile (and the third index at its median). Curves are centered to  highlight non-additive effects; changes in slope or shape indicate non-additive interaction in the statistical sense (see Figure B2 in Supplementary Material for non-centered results). Results are fairly similar across models, and there is little evidence of interaction among indices. In the constrained model, there is a suggestion of potential interaction between the first and third indices, but this evidence is untenable given the uncertainty. As more information is incorporated, not only is there less uncertainty, but the estimates appear to indicate less interaction as well. %Note that this is despite the fact that we are only incorporating prior information about the \textit{third} index, which we would expect to have modest, if any, effects on estimation of the other index curves.  
Ultimately, drawing conclusions about interactions can be difficult due to the high uncertainty. The more informative models improve these inferences by tightening credible intervals.

All models again had similar fits (RMSEs between 709 and 717), but  for this three-index model the targeted Dirichlet model with variable selection performed best (709), beating out the TEQ approach (712) and the targeted Dirichlet without variable selection (714).

\section{Discussion}
\label{s:discussion}
In this paper we have proposed several extensions to the BMIM in order to incorporate prior toxicological knowledge about multi-pollutant mixtures: constraints, transformations, and targeted informative priors based on RPFs.  A key feature of the proposed framework is that different strategies for encoding prior information may be mixed-and-matched for different indices within the same model to suit the available information (or lack thereof) about different groups of exposures. This means one could, for example, center prior weights for a multi-pollutant index around established RPFs, while imposing smoothness on weights corresponding to an exposure measured longitudinally, and leaving unstructured other exposures that are not as well understood.

Incorporating prior knowledge improves accuracy, but brings with it also the risk of incorporating incorrect information. Naturally,  approaches that treat RPFs as known \textit{a priori} (like the TEQ analysis of \citealp{mitro2016cross}) can lead to bias when those RPFs are  not transportable across studies or across populations. Even less restrictive approaches like the ordered approach described in Section \ref{ss:transformations} can perform poorly when the assumed ordering is incorrect (see Simulation B). A key benefit of the proposed informative priors is that they can improve accuracy when they encode correct prior information, but are far less sensitive to misspecification of that prior knowledge, as seen in simulations. Nevertheless, there are many ways to mis-specify a model. As has been shown for linear index models \citep{keil2020quantile}, incorrectly assuming directional homogenity can also lead to substantial bias. In Simulation C (Table A1 in Supplementary Material), we show that all the models assuming directional homogeneity (models (ii)--(vi))  exhibited increased MSE and poor interval coverage when that assumption did not hold. Incorrect assumptions about the index structure can also cause bias \citep{mcgee2021bayesian}. An area for future research is to allow for uncertainty in the index structure of a BMIM.  \cite{zavez2020modeling} proposed a latent variable model in which exposures are grouped into distinct domains, and group membership was estimated from the data. An analogous extension of the BMIM could allow for uncertainty in the index structure and could even incorporate expert knowledge via informative priors without assuming the index groupings are known.

We have proposed a novel prior specification that encodes information from  RPFs while still allowing for component-wise variable selection via spike-and-slab. Unique to this formulation is that the so-called ``slab'' is centered away from the ``spike'' (i.e., a point mass at zero). When there is very high certainty in the slab---i.e. a high concentration of mass around a positive value, and hence low mass near zero---this can lead to somewhat poor performance, with components more frequently being selected out of the model. This is unlikely in practice, however, because a high degree of certainty in a small range of values is not reasonably compatible with a 50\% probability of being zero. In the case of high-certainty in a small range of positive values, one wouldn't necessarily want to shrink estimates of that weight to zero, and one might instead opt for a Dirichlet prior without variable selection, or at least a lower prior inclusion probability.

Ultimately, the proposed methods allow one to incorporate information about relative potencies from the toxicological literature in epidemiological analyses. By the same token,  the proposed methods could be used to update and improve our understanding of the relative potencies of compounds in humans based on epidemiological data.

%\vspace{-0.5cm}
\section{Software}
\label{sec5}

Software in the form of R code is available at \url{github.com/glenmcgee/bsmim2}. Code to run simulations and data analysis is available at \url{github.com/glenmcgee/infpriormixtures}.

%\vspace{-0.5cm}
\section*{Acknowledgments}
This research was supported by NIH grants ES000002, ES028800, ES028811 and ES030990.
%{\it Conflict of Interest}: None declared.

%\bibliographystyle{imsart-nameyear} % Style BST file
%\bibliography{../../Notes/McGee_bibliography}       % Bibliography file (usually '*.bib')

\vspace{-0.5cm}
\bibliographystyle{biorefs}
\bibliography{../../Notes/McGee_bibliography}

\clearpage

\begin{table}[htbp!]
	\centering
	\caption{Simulation results across 500 datasets. Reported are mean squared error (MSE), 95\% interval width (Width), and 95\% interval coverage (Cvg; in \%). MSE and Width are reported as ratios relative to the Unconstrained model (i); values less than 1.00 indicate better performance. Holdout refers to estimated surface $h$ on a hold-out sample of 100 exposure vectors from the NHANES sample. Component-wise refers to component-wise exposure-response curves as described in Section \ref{s:MIM}, averaged over a grid of equally spaced points between the 25th and 75th percentiles. All models except for (iv) and (vi) incorporated variable selection.\label{tab:simA}}
	\begin{tabular}{ll r@{\extracolsep{5pt}} r@{\extracolsep{5pt}} r@{\extracolsep{20pt}} r@{\extracolsep{5pt}}r@{\extracolsep{5pt}}r}
		\toprule
		&& \multicolumn{3}{c}{Hold-out} & \multicolumn{3}{c}{Component-wise}  \\
		\cmidrule(l{2pt}r{2pt}){3-5}   \cmidrule(l{2pt}r{2pt}){6-8} 
		Scenario & Model &       MSE & Width & Cvg &      MSE & Width & Cvg \\ 
		\midrule
		A %$\boldsymbol{w}=\boldsymbol{w}^A$)
		& BKMR                    &  1.16 &  1.08 & 0.95 &    1.65 & 1.29 & 0.97\\ 
		& (i) Unconstrained       &  1.00 &  1.00 & 0.95 &  1.00 & 1.00 & 0.96\\ 
		& (ii) Constrained        &  0.87 &  0.96 & 0.95 & 0.80 & 0.81 & 0.95\\ 
		%& Dir-equal (No sel)     &  0.91 &  0.94 & 0.95  & 0.81 & 0.65 & 0.93\\ 
		& (iii) Dirichlet         &  0.71  &  0.89 & 0.96  & 0.54 &  0.61 & 0.96 \\
		&  (iv) Dirichlet (No Selection)&  0.65  &  0.91 & 0.97 &  0.50 &  0.65 & 0.98 \\
		& (v) Ranked             & 0.72 &  0.89 & 0.96  & 0.54 & 0.62 &0.97\\
		& (vi) TEQ                &  0.58 &  0.79 & 0.95 & 0.31 & 0.33 & 0.94\\ 
		\\[-1.8ex]   
		B %$\boldsymbol{w}=\boldsymbol{w}^B$)
		& BKMR                    & 1.17 & 1.07 & 0.95 &    1.61 & 1.28 & 0.97 \\ 
		& (i) Unconstrained       & 1.00 & 1.00 & 0.95 &  1.00 & 1.00 & 0.96 \\ 
		& (ii) Constrained        & 0.86 & 0.96 & 0.96 &  0.80 & 0.81 & 0.95 \\ 
		%& Dir-equal (No sel)     & 0.91 & 0.93 & 0.94 &  0.77 & 0.65 & 0.92 \\ 
		& (iii) Dirichlet         & 0.89 & 0.91 & 0.95  &  0.69  &  0.63 &  0.95 \\
		&  (iv) Dirichlet (No Selection)& 0.88 & 0.92 & 0.95   & 0.70  &  0.67 &  0.96 \\
		&  (v) Ranked             &1.10 &0.88 &0.92 & 0.89 & 0.57 & 0.89 \\
		& (vi) TEQ                & 2.18 & 0.81 & 0.81 &  1.55 & 0.31 & 0.61 \\ 
%		\\[-1.8ex]   
%		C  
%		& BKMR                    & 1.38 & 1.09 & 0.94 &   1.61 & 1.24 & 0.92 \\ 
%		& (i) Unconstrained       & 1.00 & 1.00 & 0.95 & 1.00 & 1.00 & 0.94 \\ 
%		& (ii) Constrained        & 1.65 & 0.94 & 0.87 &  1.76 & 0.61 & 0.68 \\ 
%		%& Dir-equal (No sel)     & 2.16 & 0.96 & 0.84 & 2.34 & 0.53 & 0.59 \\ 
%		& (iii) Dirichlet         & 1.56 & 0.88 & 0.86 &  1.60 & 0.50 & 0.61 \\
%		&  (iv) Dirichlet (No Selection)& 1.68 & 0.92 & 0.87 & 1.81 & 0.55 & 0.62 \\
%		&  (v) Ranked             & 1.72 &0.87 & 0.85  & 1.73 & 0.44 & 0.50 \\
%		& (vi) TEQ                & 2.42 & 0.84 & 0.78 & 2.47 & 0.29 & 0.24 \\ 
		\hline
	\end{tabular}
	%	\\{\scriptsize *--MSE and Width are reported as ratios relative to model (i), with values less than 1.00 indicating better performance.}
\end{table}

\end{document}